\documentclass[11pt, a4paper]{article}

\usepackage[utf8]{inputenc}
\usepackage[T1]{fontenc}
\usepackage{amsmath, amssymb, amsfonts}
\usepackage{graphicx}
\usepackage{geometry}
\usepackage{svg}
\usepackage{booktabs}
\usepackage{authblk}
\usepackage[colorlinks=true, linkcolor=blue, citecolor=blue, urlcolor=blue]{hyperref}
\usepackage{cleveref}
\usepackage{float}
\usepackage{placeins}
\usepackage[titletoc,title]{appendix}

\title{\textbf{Differentiable Multiphysics Co-Optimization via Implicit Neural Representations: A Transient Hamburger-Cooking Benchmark}}

\author[1,2,*]{Navid Zobeiry}

\affil[1]{Department of Materials Science and Engineering, University of Washington}
\affil[2]{William E. Boeing Department of Aeronautics \& Astronautics, University of Washington}
\affil[ ]{Seattle, WA 98195, USA}
\affil[*]{Corresponding author: navidz@uw.edu}

\date{\today}

\begin{document}

\maketitle

\begin{abstract}
For physical systems, the co-optimization of geometry and physical parameters, including boundary conditions (BCs), initial conditions (ICs), and material properties, remains a significant challenge. This challenge becomes more pronounced in transient multiphysics problems involving moving boundaries, strong nonlinearities, and phase transitions that introduce non-smooth changes in overall system response. Existing optimization frameworks often treat geometry and physical parameters separately or rely on simplified steady-state physics, limiting their applicability to nonlinear transient regimes relevant to design and manufacturing. Data-driven approaches, including surrogate optimization, Bayesian optimization (BO), physics-informed neural networks (PINNs), and graph neural network (GNN) surrogates, offer useful alternatives, but often require expensive offline data generation, reduced design spaces, simplified physics, or careful treatment of complex geometries, boundary conditions, and competing loss terms. In this paper, we present an end-to-end differentiable co-optimization framework that couples an implicit neural representation (INR) of geometry with a JAX-compiled Eulerian multiphysics solver. Geometry is represented as a signed distance field using Fourier-feature-encoded spatial coordinates, while physical parameters such as BCs, ICs, process controls, and material properties are optimized within the same loop. Continuous differentiable relaxations are used to represent non-smooth physical transitions while maintaining compatibility with reverse-mode automatic differentiation and Backpropagation Through Time (BPTT). 

We demonstrate the framework using a transient hamburger-cooking benchmark. This problem is selected deliberately as an interpretable multiphysics benchmark, not as a culinary optimization exercise. The benchmark contains several features that are central to difficult computational design problems: transient conductive and convective heat transfer, phase-change-driven latent energy changes, moisture and fat transport, shrinkage-induced geometry evolution, evolving contact boundary conditions, abrupt boundary-condition changes caused by flipping, and competing performance objectives. This combination makes the benchmark substantially richer than a steady-state canonical heat-transfer example, while remaining sufficiently intuitive for physical interpretation. The results show that, under fixed process conditions, the optimizer modifies geometry to relieve thermal bottlenecks. In contrast, joint co-optimization allows the design response to be distributed across geometry, material state, process variables, and boundary conditions through gradients propagated over the full transient rollout.

\end{abstract}

\section{Introduction}

Simultaneous optimization of geometry and physical parameters is central to the design of many engineered systems, including heat sinks, manufacturing processes, soft robotic systems, aerodynamic structures, and other coupled multiphysics problems. In many of these systems, the optimization problem is embedded within transient, highly nonlinear multiphysics settings, where geometry, boundary conditions (BCs), initial conditions (ICs), manufacturing parameters, and material properties are coupled through the governing physical equations and jointly influence the objective and constraint functions. In such systems, traditional sequential optimization workflows, where geometry is optimized for fixed physical parameters or physical parameters are optimized for a fixed geometry, can artificially restrict the accessible design space, bias the solution toward suboptimal designs, and miss important trade-offs arising from the coupled physics.

Classical PDE-constrained optimization has provided a central framework for addressing this class of problems. Density-based topology optimization (TO), level-set methods, and adjoint-based shape optimization have enabled major advances in structural, thermal, and fluid design \cite{bends2003topology,allaire2004structural,giles2000introduction,alexandersen2020review}. More recent studies have extended these approaches to transient multiphysics problems, such as transient heat transfer in dynamic thermal systems \cite{sun2024topography}, coupled aeroacoustics and aerodynamics \cite{monfaredi2023aeroacoustic, zhao2025conceptual}, and fluid-structure interactions subject to dynamic buckling and vortex-induced instabilities \cite{siqueira2024topology, wang2025topology}. However, implementation remains challenging when the problem involves large geometric evolution, moving interfaces, abrupt transitions, or long transient histories. In particular, transient adjoint formulations can require substantial storage or recomputation, while mesh-dependent geometry representations can complicate large topology changes and interface evolution \cite{theulings2024transient}.

Emerging data-driven methods provide a second class of approaches for reducing the need for repeated high-fidelity simulations. Surrogate-based optimization, Bayesian optimization (BO), probabilistic surrogate modeling, and multi-fidelity learning are particularly attractive when evaluations are expensive and the design space is moderately sized \cite{wang2023bo,neufang2024sbo,schoenholz2024multifidelity,fu2026peek}. Recent data-driven inverse design methods have also introduced generative models, neural parameterizations, and graph neural networks (GNNs) to represent complex geometries more directly \cite{zehnder2021ntopo,zhao2024gnnreview,gladstone2024meshgnn,hadizadeh2025gnn}. GNNs are particularly promising for mechanics and PDE-based problems because they can operate on irregular meshes and graph-structured domains, with recent studies showing strong performance in PDE prediction and shape optimization \cite{zhao2024gnnreview,gladstone2024meshgnn,hadizadeh2025gnn}. Reinforcement learning has also been explored for morphology-control and design-control co-optimization \cite{he2023morph,dai2026stackelberg}. However, these approaches are often most effective when the design space is strongly parameterized, or when large offline datasets or many environment interactions are feasible. For expensive transient multiphysics problems with evolving freeform geometry, these requirements can become limiting.

Scientific machine learning (SciML) offers a different route by integrating the underlying physics directly into the learning and optimization process. Physics-informed neural networks (PINNs) and related approaches can solve forward and inverse PDE problems without relying entirely on labeled data \cite{raissi2019physics,karniadakis2021physics}. For example, in heat-transfer and manufacturing applications, PINNs have been used to solve nonlinear heat-transfer problems and enable fast evaluation of boundary-condition effects for inverse analysis and active manufacturing control \cite{zobeiry2021pimlheat}. However, recent reviews continue to identify geometry representation, hard enforcement of BCs, and multi-objective loss balancing as central difficulties, especially in complex engineering settings \cite{plankovskyy2025pinnreview}. Differentiable simulation addresses part of this gap by retaining the numerical solver within the computational graph, allowing gradients of the physical response to be obtained directly through automatic differentiation \cite{ramsundar2021differentiable,newbury2024diffsim}. JAX-based frameworks further strengthen this direction by combining automatic differentiation, just-in-time compilation, and GPU execution, making repeated forward solves and inverse design loops more practical \cite{schoenholz2020jax,xue2023jaxfem}.

Despite these advances, a gap remains in co-optimizing freeform geometry and transient physical parameters in strongly coupled multiphysics systems. Existing approaches often treat geometry and physical parameters separately, rely on surrogate models trained on offline datasets, or restrict geometry through low-dimensional or mesh-dependent parameterizations. In addition, differentiating through long transient rollouts with moving interfaces and non-smooth physical transitions remains challenging. These limitations motivate a framework that can represent geometry continuously, couple it directly to the transient physical solver, and optimize geometry and physical parameters within the same differentiable computational graph.

To address this gap, we present an end-to-end differentiable co-optimization framework that integrates continuous geometry representation, transient multiphysics simulation, and gradient-based optimization in a single computational loop. Geometry is represented using an implicit neural representation (INR) that defines a signed distance field (SDF), enabling freeform shape evolution without explicit remeshing or a fixed low-dimensional shape parameterization. The SDF geometry and trainable physical parameters, including BCs, ICs, process controls, and material properties, are coupled directly to a differentiable Eulerian multiphysics solver implemented in JAX. Non-smooth physical mechanisms are approximated using continuous relaxations to preserve gradient flow, and the full transient response is optimized through BPTT. We demonstrate the framework using a compact but challenging hamburger-cooking benchmark involving heat and mass transport, phase change, shrinkage-driven geometry evolution, and evolving thermal contact conditions, where geometry, material composition, and cooking conditions are co-optimized. While the benchmark is intentionally interpretable, the main contribution is a general framework for joint geometry-physics optimization in transient multiphysics systems.

\section{Methodology: End-to-End Differentiable Co-Optimization}

\begin{figure}[h!]
    \centering
    \includegraphics[width=\textwidth]{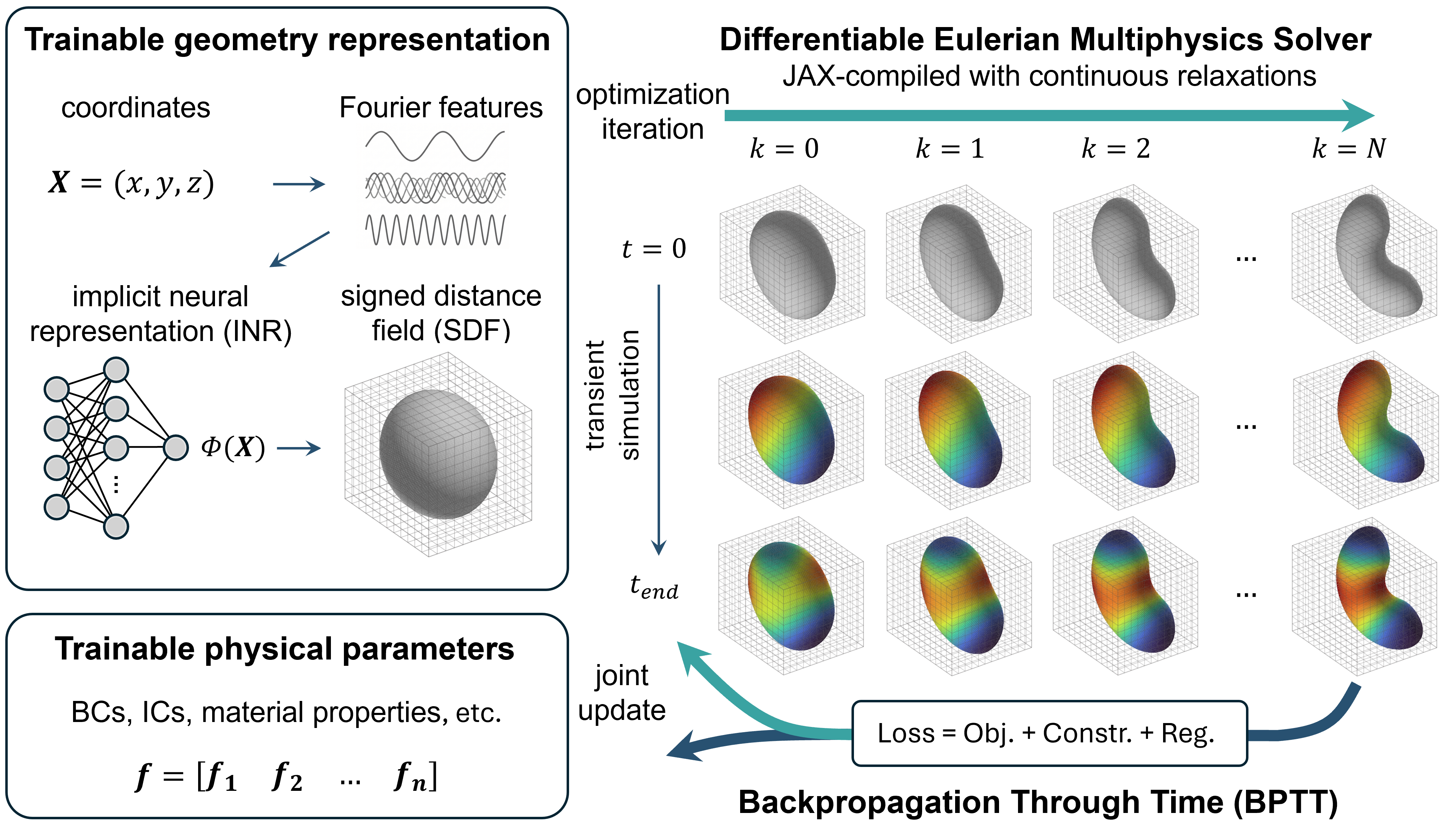}
    \caption{Overview of the differentiable co-optimization framework. Spatial coordinates are mapped through Fourier features and passed to a coordinate-based neural network to define the SDF geometry representation. The trainable geometry and physical parameters are coupled to a JAX-compiled differentiable Eulerian multiphysics solver with continuous relaxations. The transient simulation is unrolled over time, the total loss is evaluated from the solution fields, and gradients are backpropagated through the unrolled transient simulation using BPTT.}
    \label{fig:main_pipeline}
\end{figure}

The proposed framework formulates inverse design as a joint optimization problem over geometry and physical parameters within a single end-to-end differentiable transient simulation loop. As illustrated in \Cref{fig:main_pipeline}, the framework integrates a trainable coordinate-based geometry representation with a differentiable Eulerian multiphysics solver implemented in JAX. This allows gradients to propagate through the full time history of the simulation, enabling geometry and physical parameters to be updated within the same computational graph.

\subsection{Mathematical Motivation for Joint Co-Optimization}

The central motivation for the proposed framework is that geometry and physical parameters are generally coupled design variables in transient multiphysics systems. We define the optimization variables as
\begin{equation}
    \Theta = \{\Theta_{\mathrm{geo}}, \Theta_{\mathrm{phys}}\},
\end{equation}
where $\Theta_{\mathrm{geo}}$ denotes the trainable geometry parameters and $\Theta_{\mathrm{phys}}$ denotes the trainable physical parameters. Let the total objective be written as $L(\Theta_{\mathrm{geo}},\Theta_{\mathrm{phys}})$. This coupling can be expressed through the local curvature of the loss landscape,
\begin{equation}
\mathbf{H} =
\begin{bmatrix}
\nabla^2_{\Theta_{\mathrm{geo}}} L &
\nabla_{\Theta_{\mathrm{geo}}}\nabla_{\Theta_{\mathrm{phys}}} L \\
\nabla_{\Theta_{\mathrm{phys}}}\nabla_{\Theta_{\mathrm{geo}}} L &
\nabla^2_{\Theta_{\mathrm{phys}}} L
\end{bmatrix}
=
\begin{bmatrix}
\mathbf{H}_{gg} & \mathbf{H}_{gp} \\
\mathbf{H}_{pg} & \mathbf{H}_{pp}
\end{bmatrix}.
\end{equation}
The off-diagonal blocks represent cross-sensitivities between geometry and physical parameters. In transient multiphysics problems, these terms arise because the physical state evolves according to
\begin{equation}
    u^{n+1} = G(u^n,\phi,\Theta_{\mathrm{phys}};\Delta t),
\end{equation}
where $\phi$ is the implicit geometry field. When geometry and physical parameters are optimized separately, these cross-sensitivities are not explicitly used during each update. As a result, the optimizer may favor compensating strategies rather than identifying coupled geometry-physics trade-offs. The proposed differentiable formulation addresses this limitation by placing both parameter sets inside the same computational graph. The transient solver is unrolled in time, so BPTT propagates sensitivity information through the full sequence of state updates and provides coupled gradients of the form
\begin{equation}
    \nabla_{\Theta} L =
    \left[
    \nabla_{\Theta_{\mathrm{geo}}} L,
    \nabla_{\Theta_{\mathrm{phys}}} L
    \right].
\end{equation}
By keeping the full transient history differentiable, the optimizer can update geometry and physical parameters using gradients that reflect their joint effect on the objective.

\subsection{Differentiable Eulerian Multiphysics Solver (JAX-Compiled)}

The trainable geometry and physical parameters are coupled to a differentiable Eulerian multiphysics solver defined on a fixed spatial grid, as shown in \Cref{fig:main_pipeline}. In this formulation, the computational mesh remains fixed in space while the physical state variables evolve over time on that grid. This is particularly useful for transient problems involving evolving geometries and moving interfaces, since it avoids mesh distortion and remeshing challenges.

At each time step, the solver advances the physical state using the current geometry field and physical parameters. The implicit geometry field $\phi$ defines the material domain on the fixed grid, while the physical parameters define the material response, ICs, BCs, and process controls.

The full solver is implemented natively in JAX, allowing the forward simulation to remain inside a differentiable tensor program. This makes the complete transient rollout compatible with reverse-mode automatic differentiation, while just-in-time (JIT) compilation enables efficient repeated evaluation during optimization.

\subsection{Continuous Differentiable Relaxation of Non-Smooth Physics}

Transient multiphysics problems often involve sharp state transitions, such as melting, freezing, evaporation, contact changes, or threshold-based changes in material properties. If these mechanisms are implemented using discontinuous property switches or hard indicators, the forward simulation may remain numerically valid, but the gradients required for end-to-end optimization can become discontinuous, poorly conditioned, or undefined. Therefore, in the proposed framework, these transitions are represented using smooth differentiable approximations based on sigmoid gates and softplus-based thresholding. A generic relaxed indicator is written as
\begin{equation}
    \chi(z) = \sigma(\beta z) = \frac{1}{1 + e^{-\beta z}},
\end{equation}
where $\beta$ controls the sharpness of the transition. Similarly, a two-state property switch is represented as
\begin{equation}
    a(z) = a^{-}\left(1-\sigma(\beta(z-z_c))\right) + a^{+}\sigma(\beta(z-z_c)),
\end{equation}
where $z_c$ is the transition threshold. This treatment provides a differentiable numerical approximation of otherwise non-smooth mechanisms so that gradients can be propagated through the transient rollout.

\subsection{Multi-Objective Loss and Admissibility Constraints}

The proposed framework treats the design problem as a constrained multi-objective optimization problem. Since the governing solver is differentiable, objectives and constraints that depend on the transient multiphysics response can be evaluated directly from the solution fields and included in the optimization. The total loss is written as
\begin{equation}
    \mathcal{L}_{\mathrm{total}}
    =
    \mathcal{S}\left(\mathcal{L}_{1},\ldots,\mathcal{L}_{m}\right)
    +
    \mathcal{L}_{\mathrm{c}}
    +
    \mathcal{L}_{\mathrm{a}},
\end{equation}
where $\mathcal{L}_{1},\ldots,\mathcal{L}_{m}$ are scaled task-specific objective terms, $\mathcal{L}_{\mathrm{c}}$ contains problem-specific constraints, and $\mathcal{L}_{\mathrm{a}}$ contains admissibility and regularization terms. In multiphysics design problems, objectives often compete with each other. We therefore define $\mathcal{S}(\cdot)$ using a smooth Log-Sum-Exp scalarization:
\begin{equation}
    \mathcal{S}\left(\mathcal{L}_{1},\ldots,\mathcal{L}_{m}\right)
    =
    \frac{1}{\beta}
    \log
    \left[
    \sum_{i=1}^{m}
    \exp\!\left(\beta w_i \mathcal{L}_{i}\right)
    \right],
\end{equation}
where $w_i$ are additional objective weights and $\beta$ controls the sharpness of the approximation. As $\beta$ increases, this expression approaches the maximum of the weighted objectives. This form penalizes the largest remaining objective more strongly and therefore prevents the optimization from improving some objectives while neglecting the most limiting one.

The constraint term contains problem-specific requirements that are imposed as differentiable penalty terms when they are not already enforced through the design parameterization:
\begin{equation}
    \mathcal{L}_{\mathrm{c}}
    =
    \sum_{j=1}^{q}
    \lambda_j
    \mathcal{P}_j(\boldsymbol{\theta}),
\end{equation}
where $\mathcal{P}_j(\boldsymbol{\theta})$ is a nonnegative penalty associated with the violation of the $j$th constraint and $\lambda_j$ controls its contribution to the total loss. For inequality-type constraints written as $g_j(\boldsymbol{\theta}) \leq 0$, we use smooth positive-part penalties that approximate $\max(0,g_j)$, so that the penalty increases only when the constraint is violated. Scalar variables with prescribed bounds are enforced directly through bounded differentiable parameterizations and are therefore treated as hard constraints rather than soft penalty terms in the loss.

The admissibility loss includes regularization terms that keep the optimized design representation well posed during optimization. In the present implementation, admissibility is mainly associated with the implicit geometry representation. Since the geometry is represented by a signed distance field $\phi$, we include an Eikonal regularization term to preserve the signed-distance property near the interface:
\begin{equation}
    \mathcal{L}_{\mathrm{eik}}
    =
    \frac{
    \int_{\Omega_b}
    \left( \left\|\nabla \phi \right\|_2 - 1 \right)^2 \, d\mathbf{x}
    }{
    \int_{\Omega_b} d\mathbf{x}
    },
\end{equation}
where $\Omega_b$ is a narrow band around the zero level set. This term helps $\phi$ remain a valid signed distance field by encouraging $\|\nabla \phi\|_2 \approx 1$ near the boundary.

More generally, the admissibility loss is written as
\begin{equation}
    \mathcal{L}_{\mathrm{a}}
    =
    \gamma_{\mathrm{eik}}\mathcal{L}_{\mathrm{eik}}
    +
    \sum_{r=1}^{s} \gamma_r \mathcal{R}_r,
\end{equation}
where $\mathcal{R}_r$ denotes additional regularization terms. In the present study, these terms discourage poorly resolved or disconnected geometries while still allowing the optimizer to modify the shape.

\subsection{Backpropagation Through Time for Joint Co-Optimization}

As shown schematically in Figure~\ref{fig:main_pipeline}, the transient solver is unrolled in time within the differentiable computational graph. The total loss can therefore be differentiated through the full simulation history. BPTT propagates sensitivities backward through the sequence of solver updates, so the gradients account for how changes in geometry and physical parameters affect the transient response.

This is important for multiphysics design problems because performance is often determined by accumulated history rather than a single state. In this work, BPTT allows the geometry parameters and physical parameters to be updated within the same optimization loop, with gradients that reflect their coupled effect on the total loss.

\section{Case Study: Co-Optimization of Hamburger Geometry and Cooking Conditions}

In this section, we use hamburger cooking on a hot plate as a case study to demonstrate the proposed co-optimization framework. This multiphysics problem includes transient conductive and convective heat transfer, moisture and fat transport, material phase transitions such as fat rendering and liquid evaporation, shrinkage-induced geometry evolution, and evolving BCs. The hamburger is also flipped halfway through cooking, which introduces an abrupt change in the thermal BCs. The optimized variables include the 2D hamburger footprint, thickness, initial fat fraction, grill temperature, and total cooking time. In the present demonstration, the optimized 2D footprint is extruded in the thickness direction to keep the geometry representation simple and to make the flipping operation well defined. The effect of initial hamburger temperature is considered through separate cases, but the initial temperature itself is not optimized.

The hamburger is modeled as a multiphase medium composed of protein, water, and fat. The initial fat fraction is optimized as a scalar variable, while the protein, water, and fat volume fractions evolve as spatial fields during cooking, denoted by $f_p(\mathbf{x},t)$, $f_w(\mathbf{x},t)$, and $f_f(\mathbf{x},t)$.

The objective is to identify the geometry and cooking conditions that produce the desired response according to the metrics defined in this benchmark: sufficient internal heating, desirable surface browning, limited protein toughening, and high retention of water and fat. These objectives are competing. Increasing heat input or surface area can improve internal heating, but may also increase evaporation, fat loss, and surface overheating. Lower grill temperatures can reduce burning and protein toughening, but may also reduce surface browning. Changes in initial fat fraction affect heat transfer, liquid retention, and mass loss. Therefore, the optimized response cannot be determined from geometry, composition, or cooking conditions alone. It depends on their coupled effect over the full transient history.

The objective is not to develop a predictive culinary model, but to use a physically interpretable multiphysics system as a benchmark for differentiable co-optimization. It includes the main features targeted by the method: freeform geometry optimization, trainable physical parameters, transient nonlinear multiphysics, evolving boundaries, abrupt changes caused by flipping and phase transitions, evolving contact BCs, competing objectives, and differentiability through the full time history. Together, these features provide a sufficiently rich benchmark to test whether the framework can optimize geometry and physical parameters in a coupled transient setting.

\subsection{Multiphysics Hamburger Cooking Problem}

\begin{figure}[h!]
    \centering
    \includegraphics[width=0.85\textwidth]{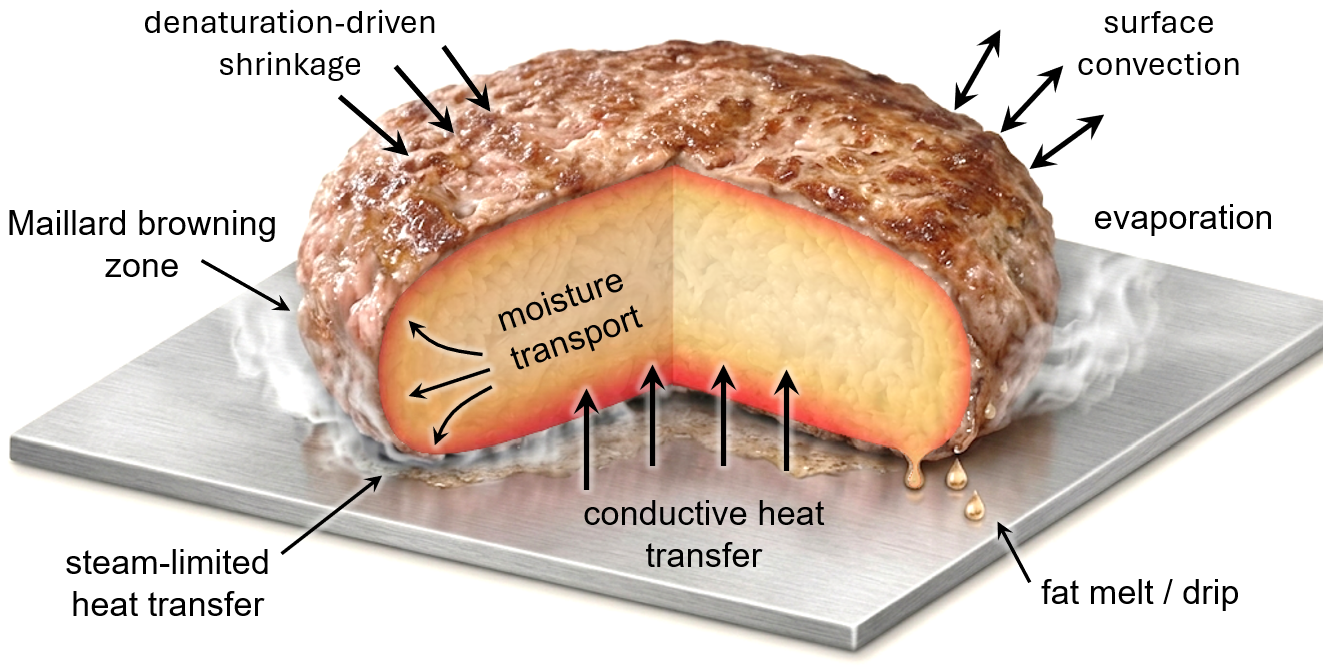}
    \caption{Schematic of the coupled heat-transfer, transport, phase-change, and shrinkage mechanisms represented in the hamburger cooking case study.}
    \label{fig:burger_physics_schematic}
\end{figure}

Figure~\ref{fig:burger_physics_schematic} summarizes the coupled physics represented in the hamburger cooking problem. Heat enters primarily through contact with the hot plate and is redistributed through the material by conduction. Exposed surfaces exchange heat with the environment. At the same time, water transport, evaporation, fat rendering and drip loss, crust formation, and protein denaturation affect the local material response. As temperature evolves, the protein matrix shrinks and the material boundary moves. Near the lower surface, contact heat transfer also changes with local wetness and steam generation.

The main state variables are the temperature field $T(\mathbf{x},t)$, the signed distance field $\phi(\mathbf{x},t)$, and the local phase fractions of water, fat, and protein. In the material domain, the temperature field is governed by
\begin{equation}
(\rho c_p)_{\mathrm{eff}}
\left(
\frac{\partial T}{\partial t}
+
\mathbf{v}\cdot\nabla T
\right)
=
\nabla \cdot
\left(
k_{\mathrm{eff}} \nabla T
\right)
-
\dot{q}_{\mathrm{vap}},
\end{equation}
where $(\rho c_p)_{\mathrm{eff}}$ is the effective volumetric heat capacity, $k_{\mathrm{eff}}$ is the effective thermal conductivity, $\mathbf{v}$ is the shrinkage-induced velocity field, and $\dot{q}_{\mathrm{vap}}$ is an effective evaporative energy sink associated with water-to-vapor transition. The hamburger is modeled as a three-phase medium composed of water, fat, and protein. These phase fractions determine the local effective thermal properties. 

Water is advected by the shrinkage-induced velocity field, redistributed through internal diffusion, and reduced by evaporation and squeeze-out
\begin{equation}
\frac{\partial f_w}{\partial t}
+
\mathbf{v}\cdot\nabla f_w
=
\nabla\cdot\left(D_{\mathrm{eff}} \nabla f_w\right)
-
\frac{\dot{S}_w}{\rho_w},
\end{equation}
where $f_w$ is the local water fraction, $D_{\mathrm{eff}}$ is the effective moisture diffusivity, and $\dot{S}_w$ is an effective water-loss term representing evaporation and squeeze-out. The effective diffusivity is reduced smoothly in hot and dry boundary regions to represent crust formation and reduced moisture transport \cite{feyissa2013roasting,dhall2012meat}. Fat is advected with the material and is lost through drip after melting, but internal fat diffusion is not included. 

Geometry evolution is driven by thermally induced shrinkage of the protein. The moving material boundary is represented by advection of the signed distance field
\begin{equation}
\frac{\partial \phi}{\partial t}
+
\mathbf{v}\cdot\nabla \phi
=
0.
\end{equation}

The BCs are divided into two types: convective heat transfer on exposed surfaces and conductive heat transfer on the hot plate. On the exposed surfaces, convection with the surrounding environment is written as
\begin{equation}
-k_{\mathrm{eff}}\nabla T \cdot \mathbf{n}
=
h_{\mathrm{env}}
\left(
T - T_{\mathrm{env}}
\right),
\qquad
\mathbf{x}\in\Gamma_{\mathrm{env}},
\end{equation}
where $\mathbf{n}$ is the outward unit normal, $h_{\mathrm{env}}$ is the convective heat transfer coefficient, and $T_{\mathrm{env}}$ is the environment temperature. Conductive BC with the hot plate is written as
\begin{equation}
-k_{\mathrm{eff}}\nabla T \cdot \mathbf{n}
=
h_{\mathrm{eff}}
\left(
T - T_{\mathrm{grill}}
\right),
\qquad
\mathbf{x}\in\Gamma_{\mathrm{plate}}.
\end{equation}
Here, $T_{\mathrm{grill}}$ is the hot-plate temperature and $h_{\mathrm{eff}}$ is an effective contact heat transfer coefficient. Unlike the environmental convection coefficient, $h_{\mathrm{eff}}$ evolves during cooking because the lower surface can transition between dry contact, wet contact, and steam-throttled contact. This is represented as
\begin{equation}
h_{\mathrm{eff}}
=
g_{\mathrm{gap}}
\left[
(1-\chi_s)
\left(
(1-\chi_w) h_{\mathrm{dry}} + \chi_w h_{\mathrm{wet}}
\right)
+
\chi_s h_{\mathrm{steam}}
\right],
\end{equation}
where $g_{\mathrm{gap}}$ accounts for distance from the plate, $\chi_w$ is a smooth wetness indicator, and $\chi_s$ is a smooth steam-barrier indicator. The hamburger is flipped once during cooking, so the surface in contact with the hot plate changes halfway through the simulation. This introduces an abrupt change in the thermal BCs.

The governing equations are solved on a fixed Eulerian grid using finite-difference operators. Simulation assumptions and constants, including material properties, BC coefficients, and phase-change thresholds, are listed in Appendix~A. Numerical details of the finite-difference implementation, including transport, surface flux treatment, flipping, and mass accounting, are summarized in Appendix~B.

\subsection{Geometry Representation and Trainable Variables}

The trainable geometry is defined by a learned 2D signed distance field $\phi_{2D}(x,y)$. In the present benchmark, this footprint is extruded in the thickness direction to define the initial 3D geometry. This keeps the top and bottom surfaces parallel to the hot plate, making the flipping operation well defined. It also isolates the effects of the in-plane footprint and thickness, without introducing additional complications associated with fully free 3D shape variations, such as gravity-driven deformation.

For the joint optimization cases, the trainable variables are
\begin{equation}
\Theta_{\mathrm{case}} =
\left\{
\Theta_{\mathrm{geo}},
T_{\mathrm{grill}},
t_{\mathrm{cook}},
t_z,
f_f^{0}
\right\},
\end{equation}
where $\Theta_{\mathrm{geo}}$ denotes the trainable parameters of the INR-based 2D geometry representation, $T_{\mathrm{grill}}$ is the hot-plate temperature, $t_{\mathrm{cook}}$ is the total cooking time, $t_z$ is the initial thickness, and $f_f^{0}$ is the initial fat fraction. The initial hamburger temperature is prescribed for each case and is not optimized. The flip time is set as a fixed fraction of the total cooking time and is not treated as an independent design variable. The scalar variables are enforced within their prescribed bounds using the bounded differentiable parameterizations described in Section~2.

\subsection{Benchmark-Specific Loss Function}

The case-study loss follows the objective-constraint-admissibility structure introduced in Section~2:
\begin{equation}
\mathcal{L}_{\mathrm{case}}
=
\mathcal{S}
\left(
\mathcal{L}_{\mathrm{core}},
\mathcal{L}_{\mathrm{brown}},
\mathcal{L}_{\mathrm{tough}},
\mathcal{L}_{\mathrm{retention}}
\right)
+
\mathcal{L}_{\mathrm{c}}^{\mathrm{case}}
+
\mathcal{L}_{\mathrm{a}}^{\mathrm{case}} .
\end{equation}
The four task objectives correspond to the benchmark definition of a desirable cooked hamburger. $\mathcal{L}_{\mathrm{core}}$ penalizes insufficient internal heating, $\mathcal{L}_{\mathrm{brown}}$ penalizes surface temperatures outside the desired browning window, $\mathcal{L}_{\mathrm{tough}}$ penalizes excessive protein toughening in the interior, and $\mathcal{L}_{\mathrm{retention}}$ penalizes loss of water and fat. These terms are scaled and aggregated using the Log-Sum-Exp scalarization defined in Section~2.

The case-specific constraint loss is
\begin{equation}
\mathcal{L}_{\mathrm{c}}^{\mathrm{case}}
=
\lambda_{\mathrm{mass}}\mathcal{L}_{\mathrm{mass}}
+
\lambda_{\mathrm{wall}}\mathcal{L}_{\mathrm{wall}}
+
\lambda_{\mathrm{center}}\mathcal{L}_{\mathrm{center}} .
\end{equation}
Here, $\mathcal{L}_{\mathrm{mass}}$ penalizes deviation from the target initial mass, $\mathcal{L}_{\mathrm{wall}}$ penalizes material outside the allowable hot-plate region, and $\mathcal{L}_{\mathrm{center}}$ penalizes drift of the footprint away from the plate center.

The admissibility loss is
\begin{equation}
\mathcal{L}_{\mathrm{a}}^{\mathrm{case}}
=
\gamma_{\mathrm{eik}}\mathcal{L}_{\mathrm{eik}}
+
\gamma_{\mathrm{top}}\mathcal{L}_{\mathrm{topology}} .
\end{equation}
The Eikonal term preserves the signed-distance character of the learned 2D footprint. The topology term discourages disconnected islands and poorly resolved fine-scale features. In the implementation, this term combines an Euler-characteristic penalty, length-scale control, and a small total-variation penalty. The corresponding parameter values and weights are listed in Appendix~A, while the implementation details are summarized in Appendix~B.

\subsection{Optimization Cases}

Two optimization modes are considered. The first is a geometry-only mode, where the hot-plate temperature, total cooking time, initial thickness, and initial fat fraction are fixed. In this mode, only the INR-based 2D geometry representation $\Theta_{\mathrm{geo}}$ is optimized. The second is a joint co-optimization mode, where $\Theta_{\mathrm{geo}}$, $t_z$, $f_f^0$, $T_{\mathrm{grill}}$, and $t_{\mathrm{cook}}$ are optimized together. This comparison separates the effect of changing only the geometry from the effect of jointly optimizing geometry, composition, and cooking conditions.

Each mode is evaluated for three prescribed initial temperatures: frozen $(-18^\circ\mathrm{C})$, refrigerated $(4^\circ\mathrm{C})$, and room temperature $(20^\circ\mathrm{C})$. The optimization cases are summarized in Table~\ref{tab:optimization_cases}.

\begin{table}[h!]
\centering
\caption{Optimization cases used in the hamburger cooking benchmark.}
\label{tab:optimization_cases}
\resizebox{\textwidth}{!}{%
\begin{tabular}{lllllll}
\toprule
Case & Mode & Optimized variables & $T_0$ & Initial $t_z$ & Initial $f_f^0$ & Initial cooking conditions \\
\midrule
1 & Geometry-only & $\Theta_{\mathrm{geo}}$ & $-18^\circ\mathrm{C}$ & $25~\mathrm{mm}$ & $0.20$ & $t_{\mathrm{cook}}=480~\mathrm{s}$, $T_{\mathrm{grill}}=220^\circ\mathrm{C}$ \\
2 & Geometry-only & $\Theta_{\mathrm{geo}}$ & $4^\circ\mathrm{C}$ & $25~\mathrm{mm}$ & $0.20$ & $t_{\mathrm{cook}}=480~\mathrm{s}$, $T_{\mathrm{grill}}=220^\circ\mathrm{C}$ \\
3 & Geometry-only & $\Theta_{\mathrm{geo}}$ & $20^\circ\mathrm{C}$ & $25~\mathrm{mm}$ & $0.20$ & $t_{\mathrm{cook}}=480~\mathrm{s}$, $T_{\mathrm{grill}}=220^\circ\mathrm{C}$ \\
4 & Joint & $\Theta_{\mathrm{geo}}, t_z, f_f^0, T_{\mathrm{grill}}, t_{\mathrm{cook}}$ & $-18^\circ\mathrm{C}$ & $25~\mathrm{mm}$ & $0.20$ & initialized at $480~\mathrm{s}$, $200^\circ\mathrm{C}$ \\
5 & Joint & $\Theta_{\mathrm{geo}}, t_z, f_f^0, T_{\mathrm{grill}}, t_{\mathrm{cook}}$ & $4^\circ\mathrm{C}$ & $25~\mathrm{mm}$ & $0.20$ & initialized at $480~\mathrm{s}$, $200^\circ\mathrm{C}$ \\
6 & Joint & $\Theta_{\mathrm{geo}}, t_z, f_f^0, T_{\mathrm{grill}}, t_{\mathrm{cook}}$ & $20^\circ\mathrm{C}$ & $25~\mathrm{mm}$ & $0.20$ & initialized at $480~\mathrm{s}$, $200^\circ\mathrm{C}$ \\
\bottomrule
\end{tabular}%
}
\end{table}

All cases use the same target initial mass of $113.4~\mathrm{g}$, corresponding to a quarter-pound hamburger, and are initialized from the same disconnected clover-like footprint. This provides a consistent and deliberately challenging starting geometry for comparing the two optimization modes. The parameter bounds, constants, and objective thresholds are listed in Appendix~A.

\section{Optimization Results}

The optimization cases defined in Table~\ref{tab:optimization_cases} are used to compare geometry-only optimization with joint co-optimization. The results are organized around two questions: how the geometry changes when cooking conditions are fixed, and how the strategy changes when geometry and cooking conditions are optimized together.

\subsection{Model Calibration}

Before running the optimization cases, selected model parameters controlling water squeeze-out, fat drip, and contact heat transfer were calibrated using a circular extruded patty. The reference calibration case used an initial thickness of $18.75~\mathrm{mm}$, initial temperature of $4^\circ\mathrm{C}$, initial fat fraction of $0.20$, initial mass of $113.4~\mathrm{g}$, total cooking time of $480~\mathrm{s}$, and grill temperature of $T_{\mathrm{grill}}=200^\circ\mathrm{C}$. This case was selected to represent a commonly reported cooking condition for a refrigerated three-quarter-inch hamburger patty.

Forward simulations were then performed by varying the selected simulation coefficients and evaluating the task-loss terms defined in Section~3.3, namely internal heating, surface browning, liquid retention, and protein toughening. The selected parameter set produced a low-loss reference response, with the interior above the target temperature, both surfaces within the browning range, moderate liquid loss, and limited protein toughening. For this case, the model predicted a mean internal maximum temperature of approximately $84.6^\circ\mathrm{C}$, mean top and bottom surface temperatures of approximately $145.1^\circ\mathrm{C}$ and $173.3^\circ\mathrm{C}$, liquid retention of approximately $71\%$, and a toughness-affected interior fraction of approximately $18.4\%$. For the initial composition used here, the $71\%$ liquid retention corresponds to approximately $23\%$ total mass loss under conserved protein mass. This calibrated parameter set was then fixed for all subsequent optimization cases. The calibrated constants are listed in Appendix~A.

\subsection{Geometry-Only Optimization}

We first examine the geometry-only optimization cases, where only the 2D footprint is trainable. The three cases use the same initial geometry and fixed cooking parameters, but differ in the initial temperature: frozen $(-18^\circ\mathrm{C})$, refrigerated $(4^\circ\mathrm{C})$, and ambient $(20^\circ\mathrm{C})$. The hot-plate temperature, total cooking time, initial thickness, and initial fat fraction are fixed at $T_{\mathrm{grill}}=220^\circ\mathrm{C}$, $t_{\mathrm{cook}}=480~\mathrm{s}$, $t_z=25~\mathrm{mm}$, and $f_f^0=0.20$, respectively. The $25~\mathrm{mm}$ thickness, which is close to one inch, was selected to create a challenging heat-transfer problem, particularly for the frozen case.

\begin{figure}[h!]
    \centering
    \includegraphics[width=\textwidth]{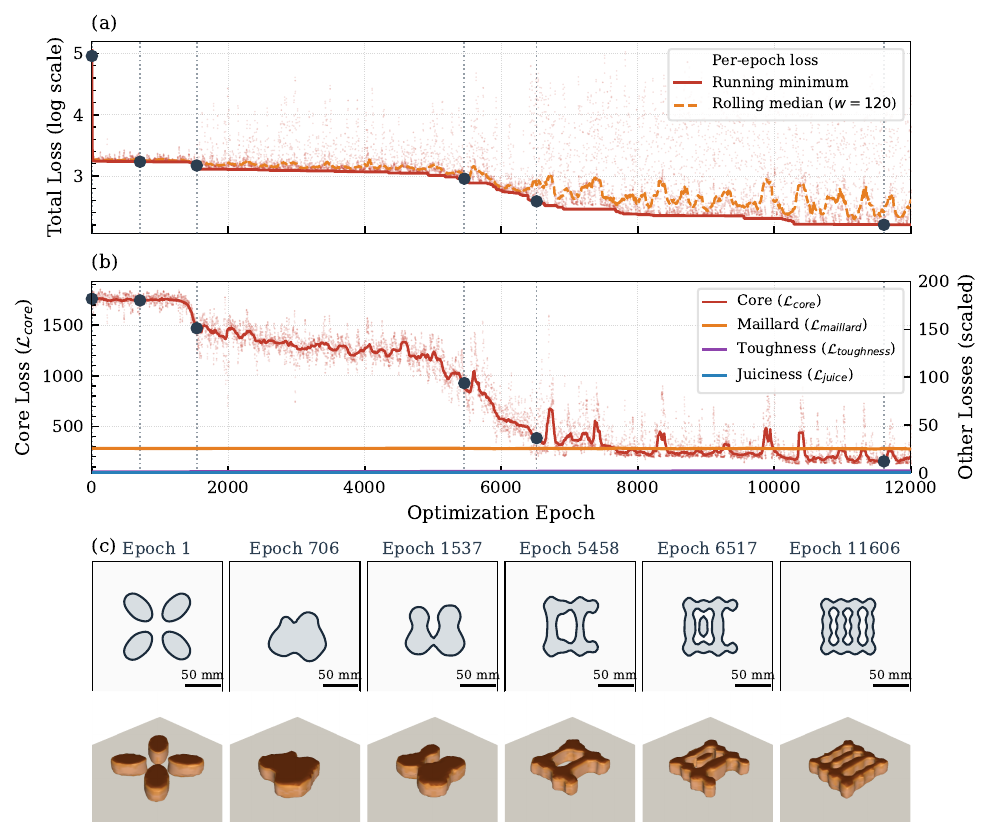} 
    \caption{Geometry-only optimization trajectory for the frozen ($-18^\circ\mathrm{C}$) case. (a) Total loss convergence, (b) task-loss evolution, and (c) evolution of the optimized 2D footprint at selected epochs.}
    \label{fig:optimization_trajectory}
\end{figure}

Figure~\ref{fig:optimization_trajectory} shows the geometry-only optimization trajectory for the frozen case. The total loss is reported on a logarithmic scale in Figure~\ref{fig:optimization_trajectory}a. The corresponding task-loss components are shown in Figure~\ref{fig:optimization_trajectory}b, and the geometry evolution is shown in Figure~\ref{fig:optimization_trajectory}c. In this case, the core-temperature loss dominates the Maillard, toughness, and liquid-retention losses. This is expected because the optimizer must raise the core temperature of a frozen, thick patty above the target internal temperature while the cooking temperature and cooking time remain fixed.

The geometry evolution shows how the optimizer responds to this constrained setting. The initial disconnected footprint first evolves toward a connected domain and then develops internal slots and a ribbed footprint. These features increase the exposed perimeter and reduce local thermal length scales, allowing the frozen case to raise the interior above the target internal temperature using geometry alone.

\begin{figure}[h!]
    \centering
    \includegraphics[width=\textwidth]{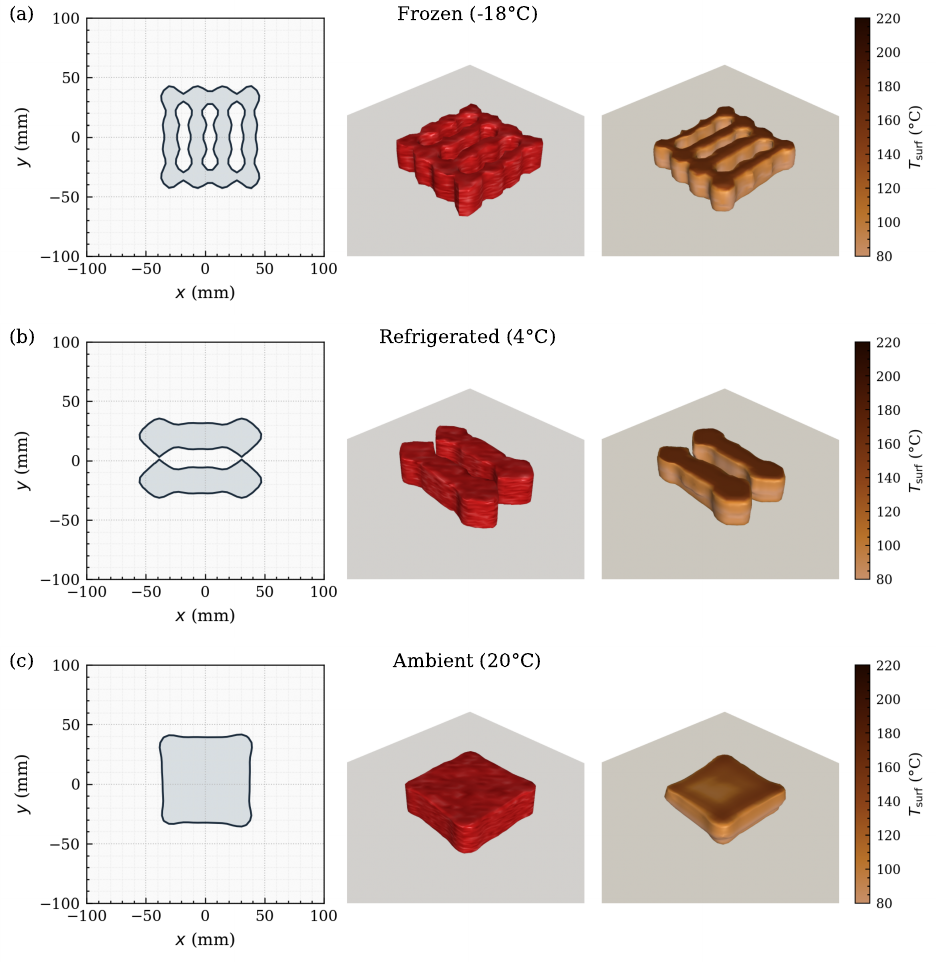} 
    \caption{Final geometry-only solutions for the three initial temperatures: (a) frozen ($-18^\circ\mathrm{C}$), (b) refrigerated ($4^\circ\mathrm{C}$), and (c) ambient ($20^\circ\mathrm{C}$).}
    \label{fig:comparative_topology}
\end{figure}

The final optimized geometries for the three initial temperatures are compared in Figure~\ref{fig:comparative_topology}. The results show a clear dependence of the optimized geometry on the initial thermal state. The frozen case produces the largest geometric change, with multiple narrow slots. The refrigerated case produces two elongated lobes separated by a central gap, increasing surface exposure without forming the strongly slotted geometry observed in the frozen case. The ambient case converges to a compact rounded-square footprint, which is sufficient for internal heating while reducing unnecessary surface exposure and liquid loss.

\subsection{Joint Optimization of Geometry, Composition, and Cooking Conditions}

\begin{figure}[h!]
    \centering
    \includegraphics[width=\textwidth]{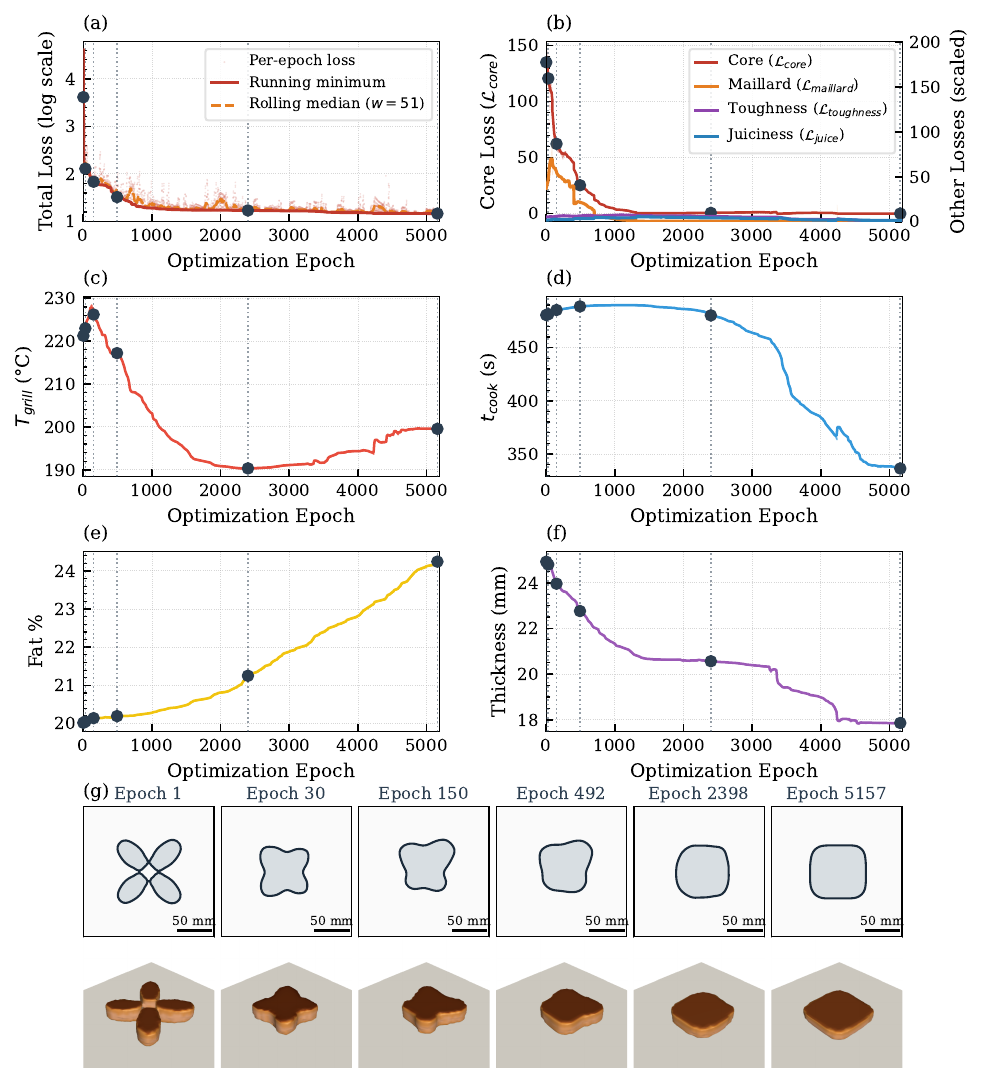} 
    \caption{Joint optimization trajectory for the refrigerated ($4^\circ\mathrm{C}$) case. (a) Total loss convergence, (b) task-loss evolution, (c) optimized hot-plate temperature, (d) optimized cooking time, (e) optimized initial fat fraction, (f) optimized thickness, and (g) evolution of the optimized geometry at selected epochs.}
    \label{fig:joint_optimization}
\end{figure}

We next examine the joint optimization cases, where the 2D geometry, thickness, fat fraction, hot-plate temperature, and total cooking time are optimized together. Figure~\ref{fig:joint_optimization} shows the optimization trajectory for the refrigerated case, $T_0=4^\circ\mathrm{C}$.

The early stage of optimization, approximately the first $1000$ epochs, is dominated by the core-temperature and surface-browning losses (Figure~\ref{fig:joint_optimization}b). During this stage, the optimizer reduces these losses primarily through the scalar physical variables. The thickness decreases, which shortens the thermal length scale and improves internal heating (Figure~\ref{fig:joint_optimization}f). At the same time, the hot-plate temperature decreases, reducing excessive surface heating and helping control the browning loss (Figure~\ref{fig:joint_optimization}c). The 2D footprint changes more modestly during this stage, mainly by merging the initially disconnected clover-like regions into a connected domain to satisfy the admissibility constraints (Figure~\ref{fig:joint_optimization}g).

After approximately $1000$ epochs, the optimization enters a slower refinement stage. The total loss continues to decrease through coupled adjustments of cooking time, fat fraction, thickness, and geometry. The cooking time decreases substantially from its initial value of $480~\mathrm{s}$ to about $336~\mathrm{s}$ (Figure~\ref{fig:joint_optimization}d), while the fat fraction increases from $20\%$ to about $24\%$ (Figure~\ref{fig:joint_optimization}e). In parallel, the footprint gradually evolves toward a compact rounded-square shape. 

The same qualitative outcome is observed across the three joint optimization cases. In all cases, the optimized geometry converges to a compact rounded-square footprint with a lower final thickness, while fat fraction, hot-plate temperature, and cooking time adapt to the initial thermal state. Table~\ref{tab:joint_final_results} summarizes the final joint optimization results.

\begin{table}[h!]
\centering
\caption{Final values from the joint optimization cases.}
\label{tab:joint_final_results}
\begin{tabular}{llllll}
\toprule
Initial condition & Final $t_z$ & Final $f_f^0$ & Final $t_{\mathrm{cook}}$ & Final $T_{\mathrm{grill}}$ & Final geometry \\
\midrule
Frozen ($-18^\circ\mathrm{C}$) & $18.0~\mathrm{mm}$ & $21.1\%$ & $491~\mathrm{s}$ & $192^\circ\mathrm{C}$ & Rounded square \\
Refrigerated ($4^\circ\mathrm{C}$) & $17.9~\mathrm{mm}$ & $24.2\%$ & $336~\mathrm{s}$ & $200^\circ\mathrm{C}$ & Rounded square \\
Ambient ($20^\circ\mathrm{C}$) & $21.5~\mathrm{mm}$ & $22.1\%$ & $349~\mathrm{s}$ & $197^\circ\mathrm{C}$ & Rounded square \\
\bottomrule
\end{tabular}
\end{table}

\subsection{Discussion and Limitations}

The benchmark studied here contains several features that are relevant to broader transient multiphysics design problems: evolving geometry, moving material boundaries, phase-dependent properties, abrupt BC changes, competing objectives, and history-dependent performance metrics. Therefore, it provides a sufficiently complex demonstration problem for evaluating whether the proposed framework can jointly optimize geometry and physical parameters in a transient multiphysics setting.

The results highlight the difference between geometry-only optimization and geometry-physics co-optimization. In the geometry-only cases, the optimized footprint changes strongly with the initial temperature. This effect is most visible for the frozen thick-patty case, where the optimizer forms a high-perimeter, slotted footprint. This geometry reduces the effective thermal length scale, allowing the core temperature to exceed the target internal temperature despite fixed cooking time, fixed thickness, and fixed hot-plate temperature.

For the joint optimization cases, the optimizer first uses the scalar physical variables, particularly thickness and hot-plate temperature, to control the through-thickness heat-transfer problem. Once the dominant thermal losses are reduced, the geometry evolves toward a compact rounded-square footprint. This geometry avoids unnecessary surface exposure while still providing sufficient heat input through the exposed boundaries. The result is a more balanced solution in which geometry, thickness, temperature, cooking time, and composition act together to satisfy the competing heating, browning, toughness, and liquid-retention objectives.

These results support the central motivation of the framework. When geometry and physical parameters are optimized in the same differentiable rollout, the optimizer can distribute the design response across shape, composition, and cooking conditions rather than forcing geometry alone to compensate for fixed operating conditions. However, because the benchmark uses calibrated cooking coefficients, prescribed loss thresholds, and an extruded 2D geometry representation, the results should be interpreted as a specific computational demonstration rather than a general validation of the framework. Even within this restricted setting, the results show that the method can co-optimize geometry and physical parameters in a transient multiphysics problem.

\section{Ablation Analysis of Coupled Design Variables}

The results in Section 4 suggest that the full joint optimizer uses thickness as an important design variable to reduce the dominant through-thickness thermal bottleneck. To examine this effect more directly, we performed an additional refrigerated-case ablation in which the initial thickness was fixed at its initial value of 25 mm, while the in-plane geometry, initial fat fraction, hot-plate temperature, and cooking time remained trainable. All other settings were kept the same as the refrigerated joint optimization case.

This case is compared with the full joint refrigerated case and the geometry-only refrigerated case in Table~\ref{tab:thickness_ablation}. The geometry-only case is included as a mechanistic reference rather than a strict quantitative control, since it was run with fixed cooking conditions of $T_{\mathrm{grill}}=220^\circ$C and $t_{\mathrm{cook}}=480$ s. In contrast, the fixed-thickness joint case allows $T_{\mathrm{grill}}$, $t_{\mathrm{cook}}$, and $f_f^0$ to evolve. Nevertheless, both cases retain the same fixed initial thickness of 25 mm, and both produce elongated two-lobed footprints rather than the compact rounded-square geometry obtained in the full joint case.

\begin{table}[h]
\centering
\caption{Refrigerated-case ablation used to examine the role of thickness as a coupled design variable.}
\label{tab:thickness_ablation}
\begin{tabular}{lccccc}
\hline
Case & Trainable thickness & $t_z$ & $f_f^0$ & $t_{\mathrm{cook}}$ & $T_{\mathrm{grill}}$ \\
\hline
Geometry-only & No & 25.0 mm & 20.0\% & 480 s & 220$^\circ$C \\
Joint, fixed thickness & No & 25.0 mm & 25.9\% & 500 s & 206$^\circ$C \\
Full joint & Yes & 17.9 mm & 24.2\% & 336 s & 200$^\circ$C \\
\hline
\end{tabular}
\end{table}

The fixed-thickness ablation shows that, when the optimizer cannot reduce the through-thickness thermal length scale directly, it compensates through the remaining design variables. The cooking time is driven close to the upper bound, the fat fraction increases, and the in-plane footprint evolves toward two elongated regions. This footprint is qualitatively similar to the refrigerated geometry-only solution, indicating that in-plane geometry becomes a compensating mechanism when thickness is unavailable.

In contrast, the full joint case reaches a more balanced strategy. By reducing the initial thickness, the optimizer directly reduces the through-thickness heat-transfer length scale, allowing the cooking time to decrease substantially and the in-plane geometry to remain more compact, forming a rounded-rectangular footprint. This result supports the main conclusion that co-optimization does not simply improve the objective by adding more variables, but by allowing the optimizer to distribute the design response across dominant variables that control different bottlenecks in the transient multiphysics response.

\section{Conclusion}

This paper presented an end-to-end differentiable framework for joint optimization of geometry and physical parameters in transient multiphysics problems. The framework combines an INR-based SDF geometry representation with a JAX-compiled Eulerian solver, allowing geometry, BCs, ICs, material parameters, and process variables to be optimized within the same computational graph. Smooth differentiable relaxations were used to represent threshold-based physical transitions, and BPTT was used to propagate gradients through the full transient rollout.

The hamburger cooking benchmark was used as a compact demonstration problem involving transient conductive and convective heat transfer, moisture and fat transport, phase change, shrinkage-induced geometry evolution, evolving contact BCs, and competing performance objectives. The geometry-only cases showed that, when cooking conditions are fixed, the optimizer modifies the footprint strongly in response to the initial temperature. For the frozen thick-patty case, this led to a high-perimeter slotted geometry that improved internal heating under fixed time, thickness, and temperature constraints.

The joint optimization cases showed a different behavior. When geometry, thickness, fat fraction, hot-plate temperature, and cooking time were optimized together, the optimizer first reduced the loss by modifying the thickness and hot-plate temperature, which directly control the dominant through-thickness heat-transfer response. The geometry then evolved toward a compact rounded-square footprint while the remaining scalar variables were optimized simultaneously. In the geometry-only cases, thickness and hot-plate temperature are fixed, so the optimizer must compensate through the 2D footprint. In the joint cases, the optimizer can instead use the variables that most directly control the dominant physics and distribute the design response across geometry, composition, temperature, and time.

The fixed-thickness ablation further clarified the role of coupled design variables. When the initial thickness was held fixed, the optimizer compensated by increasing cooking time, increasing fat fraction, and modifying the in-plane footprint toward an elongated two-lobed geometry. In contrast, the full joint case reduced the thickness directly and produced a more compact rounded-rectangular footprint with a substantially shorter cooking time. This confirms that the benefit of co-optimization is not simply the addition of more variables, but the ability to use physically meaningful variables that control different bottlenecks in the transient response.

Overall, the results show that the proposed differentiable formulation can use coupled gradients through the full transient history to optimize both geometry and physical parameters. This is the main contribution of the work: the design response is not limited to shape modification, but can be distributed across the variables that control the governing physics. This provides a practical route for inverse design problems where geometry, material state, BCs, and process variables are strongly coupled.

\bibliographystyle{unsrt} 
\bibliography{references} 

\begin{thebibliography}{10}

\bibitem{bends2003topology}
Martin~Philip Bends{\o}e and Ole Sigmund.
\newblock {\em Topology Optimization: Theory, Methods, and Applications}.
\newblock Springer, Berlin, Heidelberg, 2 edition, 2004.

\bibitem{allaire2004structural}
Gr{\'e}goire Allaire, Fran{\c{c}}ois Jouve, and Anca-Maria Toader.
\newblock Structural optimization using sensitivity analysis and a level-set method.
\newblock {\em Journal of Computational Physics}, 194(1):363--393, 2004.

\bibitem{giles2000introduction}
Michael~B. Giles and Niles~A. Pierce.
\newblock An introduction to the adjoint approach to design.
\newblock {\em Flow, Turbulence and Combustion}, 65(3--4):393--415, 2000.

\bibitem{alexandersen2020review}
Joe Alexandersen and Casper~Schousboe Andreasen.
\newblock A review of topology optimisation for fluid-based problems.
\newblock {\em Fluids}, 5(1):29, 2020.

\bibitem{sun2024topography}
Yupeng Sun, Song Yao, and Joe Alexandersen.
\newblock Topography optimisation using a reduced-dimensional model for transient conjugate heat transfer between fluid channels and solid plates with volumetric heat source.
\newblock {\em Structural and Multidisciplinary Optimization}, 67(4):45, 2024.

\bibitem{monfaredi2023aeroacoustic}
Morteza Monfaredi, Varvara Asouti, Xenofon Trompoukis, Konstantinos Tsiakas, and Kyriakos Giannakoglou.
\newblock Aeroacoustic and aerodynamic adjoint-based shape optimization of an axisymmetric aero-engine intake.
\newblock {\em Aerospace}, 10(9):743, 2023.

\bibitem{zhao2025conceptual}
Peng Liao, Peng Bai, Yudong Zhang, and Chuanzhen Liu.
\newblock Conceptual design of supersonic {Busemann} biplane based on neighborhood-search aerodynamic topology.
\newblock {\em Physics of Fluids}, 37(3):031702, 2025.

\bibitem{siqueira2024topology}
Lucas~Oliveira Siqueira, Kamilla Emily~Santos Silva, Em{\'i}lio Carlos~Nelli Silva, and Renato~Picelli Sanches.
\newblock Topology optimization of stationary fluid--structure interaction problems considering a natural frequency constraint for vortex-induced vibrations attenuation.
\newblock {\em Finite Elements in Analysis and Design}, 234:104137, 2024.

\bibitem{wang2025topology}
Qingdi Wang, Lucas~Oliveira Siqueira, Tao Xu, Min Zhao, Renato Picelli, and Yi~Min Xie.
\newblock Topology optimization of fluid--structure interaction problems with buckling constraints.
\newblock {\em Computers \& Structures}, 318:107978, 2025.

\bibitem{theulings2024transient}
M.~J.~B. Theulings, R.~Maas, Lisa No{\"e}l, Fred van Keulen, and Matthijs Langelaar.
\newblock Reducing time and memory requirements in topology optimization of transient problems.
\newblock {\em International Journal for Numerical Methods in Engineering}, 125(14):e7461, 2024.

\bibitem{wang2023bo}
Xilu Wang, Yaochu Jin, Sebastian Schmitt, and Markus Olhofer.
\newblock Recent advances in bayesian optimization.
\newblock {\em ACM Computing Surveys}, 55(13s):1--36, 2023.

\bibitem{neufang2024sbo}
Mathias Neufang, Emma Pajak, Damien van~de Berg, Ye~Seol Lee, and Ehecatl~Antonio Del Rio~Chanona.
\newblock Surrogate-based optimization techniques for process systems engineering.
\newblock In Fernando~Israel G{\'o}mez-Castro and Vicente Rico-Ram{\'i}rez, editors, {\em Optimization in Chemical Engineering: Deterministic, Meta-Heuristic and Data-Driven Techniques}, pages 159--214. De Gruyter, Berlin, Boston, 2025.

\bibitem{schoenholz2024multifidelity}
Caleb Schoenholz, Enrico Zappino, Marco Petrolo, and Navid Zobeiry.
\newblock Efficient analysis of composites manufacturing using multi-fidelity simulation and probabilistic machine learning.
\newblock {\em Composites Part B: Engineering}, 280:111499, 2024.

\bibitem{fu2026peek}
Huilong Fu and Navid Zobeiry.
\newblock Data-driven machine learning meta-analysis of process--property relationships in polymer additive manufacturing: A case study on {FFF}-printed {PEEK}.
\newblock {\em Journal of Manufacturing Processes}, 163:100--113, 2026.

\bibitem{zehnder2021ntopo}
Jonas Zehnder, Yue Li, Stelian Coros, and Bernhard Thomaszewski.
\newblock {NTopo}: Mesh-free topology optimization using implicit neural representations.
\newblock In {\em Advances in Neural Information Processing Systems}, volume~34, pages 10368--10381, 2021.

\bibitem{zhao2024gnnreview}
Yingxue Zhao, Haoran Li, Haosu Zhou, Hamid~Reza Attar, Tobias Pfaff, and Nan Li.
\newblock A review of graph neural network applications in mechanics-related domains.
\newblock {\em Artificial Intelligence Review}, 57:315, 2024.

\bibitem{gladstone2024meshgnn}
Rini~Jasmine Gladstone, Helia Rahmani, Vishvas Suryakumar, Hadi Meidani, Marta D'Elia, and Ahmad Zareei.
\newblock Mesh-based {GNN} surrogates for time-independent {PDEs}.
\newblock {\em Scientific Reports}, 14(1):3394, 2024.

\bibitem{hadizadeh2025gnn}
Farnoosh Hadizadeh, Wrik Mallik, and Rajeev~K. Jaiman.
\newblock A graph neural network surrogate model for multi-objective fluid-acoustic shape optimization.
\newblock {\em Computer Methods in Applied Mechanics and Engineering}, 441:117921, 2025.

\bibitem{he2023morph}
Zhanpeng He and Matei Ciocarlie.
\newblock {MORPH}: Design co-optimization with reinforcement learning via a differentiable hardware model proxy.
\newblock In {\em Proceedings of the IEEE International Conference on Robotics and Automation}, pages 7764--7771, 2024.

\bibitem{dai2026stackelberg}
Tianjian Chen, Zhanpeng He, and Matei Ciocarlie.
\newblock Co-designing hardware and control for robot hands.
\newblock {\em Science Robotics}, 6(54):eabg2133, 2021.

\bibitem{raissi2019physics}
Maziar Raissi, Paris Perdikaris, and George~Em Karniadakis.
\newblock Physics-informed neural networks: A deep learning framework for solving forward and inverse problems involving nonlinear partial differential equations.
\newblock {\em Journal of Computational Physics}, 378:686--707, 2019.

\bibitem{karniadakis2021physics}
George~Em Karniadakis, Ioannis~G. Kevrekidis, Lu~Lu, Paris Perdikaris, Sifan Wang, and Liu Yang.
\newblock Physics-informed machine learning.
\newblock {\em Nature Reviews Physics}, 3(6):422--440, 2021.

\bibitem{zobeiry2021pimlheat}
Navid Zobeiry and Keith~D. Humfeld.
\newblock A physics-informed machine learning approach for solving heat transfer equation in advanced manufacturing and engineering applications.
\newblock {\em Engineering Applications of Artificial Intelligence}, 101:104232, 2021.

\bibitem{plankovskyy2025pinnreview}
Sergiy Plankovskyy, Yevgen Tsegelnyk, Nataliia Shyshko, Igor Litvinchev, Tetyana Romanova, and Jos{\'e}~Manuel Velarde~Cant{\'u}.
\newblock Review of physics-informed neural networks: Challenges in loss function design and geometric integration.
\newblock {\em Mathematics}, 13(20):3289, 2025.

\bibitem{ramsundar2021differentiable}
Yuanming Hu, Luke Anderson, Tzu-Mao Li, Qi~Sun, Nathan Carr, Jonathan Ragan-Kelley, and Fr{\'e}do Durand.
\newblock {DiffTaichi}: Differentiable programming for physical simulation.
\newblock In {\em International Conference on Learning Representations}, 2020.

\bibitem{newbury2024diffsim}
Rhys Newbury, Jack Collins, Kerry He, Jiahe Pan, Ingmar Posner, David Howard, and Akansel Cosgun.
\newblock A review of differentiable simulators.
\newblock {\em IEEE Access}, 12:97581--97604, 2024.

\bibitem{schoenholz2020jax}
Samuel~S. Schoenholz and Ekin~D. Cubuk.
\newblock {JAX}, {MD}: A framework for differentiable physics.
\newblock In {\em Advances in Neural Information Processing Systems}, volume~33, pages 11428--11441, 2020.

\bibitem{xue2023jaxfem}
Tianju Xue, Shuheng Liao, Zhengtao Gan, Chanwook Park, Xiaoyu Xie, Wing~Kam Liu, and Jian Cao.
\newblock {JAX-FEM}: A differentiable {GPU}-accelerated 3d finite element solver for automatic inverse design and mechanistic data science.
\newblock {\em Computer Physics Communications}, 291:108802, 2023.

\bibitem{feyissa2013roasting}
Aberham~Hailu Feyissa, Krist~V. Gernaey, and Jens Adler-Nissen.
\newblock 3d modelling of coupled mass and heat transfer of a convection-oven roasting process.
\newblock {\em Meat Science}, 93(4):810--820, 2013.

\bibitem{dhall2012meat}
Ashish Dhall, Amit Halder, and Ashim~K. Datta.
\newblock Multiphase and multicomponent transport with phase change during meat cooking.
\newblock {\em Journal of Food Engineering}, 113:299--309, 2012.

\end{thebibliography}

\clearpage
\begin{appendices}

\section{Benchmark Assumptions}

\subsection{Initial Composition}

The hamburger is initialized as a three-phase mixture of water, fat, and protein. The initial fat fraction $f_f^0$ is prescribed or optimized depending on the case. The remaining lean fraction is partitioned as
\begin{equation}
    f_w^0 = 0.75(1-f_f^0),
    \qquad
    f_p^0 = 0.25(1-f_f^0).
\end{equation}

\subsection{Material Constants and Thermal Property Models}

\begin{table}[H]
\centering
\caption{Material constants.}
\label{tab:material_constants}
\begin{tabular}{lll}
\toprule
Quantity & Value & Description \\
\midrule
$\rho_w$ & $1000~\mathrm{kg/m^3}$ & Water density \\
$\rho_f$ & $900~\mathrm{kg/m^3}$ & Fat density \\
$\rho_p$ & $1050~\mathrm{kg/m^3}$ & Protein density \\
$H_{\mathrm{vap}}$ & $2.26\times10^6~\mathrm{J/kg}$ & Latent heat of vaporization \\
$L_{\mathrm{fusion}}$ & $3.34\times10^5~\mathrm{J/kg}$ & Latent heat of fusion \\
$\Delta T_{\mathrm{melt}}$ & $15^\circ\mathrm{C}$ & Apparent heat-capacity smoothing width \\
$T_{\mathrm{melt}}$ & $57^\circ\mathrm{C}$ & Fat melting threshold \\
$T_{\mathrm{denat}}$ & $65^\circ\mathrm{C}$ & Protein denaturation threshold \\
\bottomrule
\end{tabular}
\end{table}

\begin{table}[H]
\centering
\caption{Temperature-dependent thermal property models. Temperature is in $^\circ\mathrm{C}$.}
\label{tab:thermal_property_models}
\begin{tabular}{ll}
\toprule
Quantity & Model \\
\midrule
$k_w$ & $0.56 + 0.0018(T-20)$ \\
$k_f$ & $\max\left(0.05,\;0.20-0.0005\,\mathrm{ReLU}(T-20)\right)$ \\
$k_p$ & $0.40+0.0010\,\mathrm{ReLU}(T-20)$ \\
$c_{p,f}$ & $2000+5\,\mathrm{ReLU}(T-20)$ \\
$c_{p,p}$ & $2000+2\,\mathrm{ReLU}(T-T_{\mathrm{denat}})$ \\
$c_{p,w}$ &
$c_{p,\mathrm{base}}+
\dfrac{L_{\mathrm{fusion}}}{\Delta T_{\mathrm{melt}}\sqrt{\pi}}
\exp\left[-\left(\dfrac{T}{\Delta T_{\mathrm{melt}}}\right)^2\right]$ \\
\bottomrule
\end{tabular}
\end{table}

\subsection{Boundary-Condition, Contact, and Transport Constants}

\begin{table}[H]
\centering
\caption{Thermal BC, contact, transport, and calibrated loss constants.}
\label{tab:bc_transport_constants}
\begin{tabular}{lll}
\toprule
Quantity & Value & Description \\
\midrule
$T_{\mathrm{env}}$ & $20^\circ\mathrm{C}$ & Environment temperature \\
$h_{\mathrm{env}}$ & $35~\mathrm{W/m^2K}$ & Environmental convection coefficient \\
$h_{\mathrm{dry}}$ & $340~\mathrm{W/m^2K}$ & Dry contact coefficient \\
$h_{\mathrm{wet}}$ & $650~\mathrm{W/m^2K}$ & Wet contact coefficient \\
$h_{\mathrm{steam}}$ & $150~\mathrm{W/m^2K}$ & Steam-throttled contact coefficient \\
$\ell_{\mathrm{gap}}$ & $0.025~\mathrm{m}$ & Contact decay length \\
$q_{\mathrm{steam}}^\ast$ & $2.5\times10^4~\mathrm{W/m^2}$ & Steam-barrier flux threshold \\
$\Delta q_{\mathrm{steam}}$ & $6.0\times10^3~\mathrm{W/m^2}$ & Steam-barrier transition width \\
$f_{w,\mathrm{wet}}$ & $0.02$ & Wet-contact water threshold \\
$\alpha_w$ & $250$ & Wet-contact sharpness \\
$T_{\mathrm{boil}}$ & $95^\circ\mathrm{C}$ & Boiling switch temperature \\
$\alpha_{\mathrm{boil}}$ & $2$ & Boiling switch sharpness \\
$D_0$ & $1.0\times10^{-9}~\mathrm{m^2/s}$ & Baseline moisture diffusivity \\
$c_{\mathrm{sqz}}$ & $6.0\times10^{-4}$ & Water squeeze-out coefficient \\
$c_{\mathrm{fat}}$ & $2.25\times10^{-3}$ & Fat-drip coefficient \\
$f_{w,\min}$ & $0.05$ & Minimum water fraction for evaporation \\
$T_{\mathrm{env,local}}$ &
$T_{\mathrm{env}}+150\exp(-d_{\mathrm{plate}}/0.030)$ &
Local air temperature near plate \\
\bottomrule
\end{tabular}
\end{table}

\subsection{Optimization Bounds and Initial Values}

\begin{table}[H]
\centering
\caption{Bounds and initial scalar values used in the optimization cases.}
\label{tab:optimization_bounds_initial_values}
\begin{tabular}{llll}
\toprule
Quantity & Bounds & Geometry-only value & Joint initial value \\
\midrule
$T_{\mathrm{grill}}$ & $150$--$300^\circ\mathrm{C}$ & $220^\circ\mathrm{C}$ & $200^\circ\mathrm{C}$ \\
$t_{\mathrm{cook}}$ & $20$--$500~\mathrm{s}$ & $480~\mathrm{s}$ & $480~\mathrm{s}$ \\
$t_z$ & $8$--$28~\mathrm{mm}$ & $25~\mathrm{mm}$ & $25~\mathrm{mm}$ \\
$f_f^0$ & $0.15$--$0.30$ & $0.20$ & $0.20$ \\
$m_0$ & -- & $113.4~\mathrm{g}$ & $113.4~\mathrm{g}$ \\
\bottomrule
\end{tabular}
\end{table}

One flip is applied at
\begin{equation}
    t_{\mathrm{flip}}=0.5t_{\mathrm{cook}}.
\end{equation}

\subsection{Objective Thresholds}

\begin{table}[H]
\centering
\caption{Objective thresholds.}
\label{tab:objective_thresholds}
\begin{tabular}{lll}
\toprule
Quantity & Value & Description \\
\midrule
$T_{\mathrm{core}}^\ast$ & $68^\circ\mathrm{C}$ & Target internal temperature \\
$T_{\mathrm{tough}}$ & $75^\circ\mathrm{C}$ & Toughness-exposure threshold \\
$T_{\mathrm{brown,low}}$ & $140^\circ\mathrm{C}$ & Lower browning threshold \\
$T_{\mathrm{brown,high}}$ & $195^\circ\mathrm{C}$ & Upper browning threshold \\
$T_{\mathrm{brown,ideal}}$ & $167.5^\circ\mathrm{C}$ & Browning midpoint \\
$r_{\mathrm{ret}}^\ast$ & $0.69$ & Target liquid-retention ratio \\
$r_{\mathrm{ret,danger}}$ & $0.60$ & Strong retention-penalty threshold \\
$\delta_{\mathrm{core}}$ & $2.5~\mathrm{mm}$ & Interior-mask gray-band thickness \\
\bottomrule
\end{tabular}
\end{table}

\FloatBarrier

\section{Finite-Difference and Optimization Implementation}

\subsection{Grid, Time Integration, and Rollout}

\begin{table}[H]
\centering
\caption{Grid and time-integration settings.}
\label{tab:grid_time_settings}
\begin{tabular}{lll}
\toprule
Quantity & Value & Description \\
\midrule
Grid resolution & $50\times50\times50$ & Cartesian grid \\
Domain size & $0.20\times0.20\times0.06~\mathrm{m}$ & $x$, $y$, and $z$ extents \\
$\Delta t$ & $2~\mathrm{s}$ & Nominal time step \\
$N_{\mathrm{steps}}$ & $250$ & Maximum rollout length \\
Time mask & $\Delta t_n=\Delta t\,\sigma(t_{\mathrm{cook}}-t_n)$ & Smooth end-of-cook masking \\
Rollout & \texttt{jax.lax.scan} & Differentiable time unrolling \\
Memory control & Checkpointing inside scan & Reduced reverse-mode storage \\
\bottomrule
\end{tabular}
\end{table}

\subsection{Geometry Network and Initialization}

\begin{table}[H]
\centering
\caption{INR geometry model.}
\label{tab:inr_model}
\begin{tabular}{lll}
\toprule
Quantity & Value & Description \\
\midrule
Input & $(x,y)$ & 2D footprint coordinates \\
Encoding & Fourier features & $\sin(f_i x)/f_i$, $\cos(f_i x)/f_i$, and corresponding $y$ terms \\
Frequencies & $f_i=1.5^i,\; i=0,\ldots,11$ & Logarithmic spacing \\
MLP depth & $4$ hidden layers & Geometry network \\
MLP width & $64$ hidden features & Per hidden layer \\
Activation & Swish & Hidden layers \\
Output & $\phi_{2D}(x,y)$ & Learned 2D SDF footprint \\
\bottomrule
\end{tabular}
\end{table}

The 3D SDF is constructed by extrusion:
\begin{equation}
    \phi(x,y,z)
    =
    \max
    \left[
    \phi_{2D}(x,y),
    \left|z-z_c(t_z)\right|-\frac{t_z}{2}
    \right],
\end{equation}
where $z_c(t_z)$ places the lower surface on the hot plate.

\begin{table}[H]
\centering
\caption{Geometry initialization.}
\label{tab:geometry_initialization}
\begin{tabular}{lll}
\toprule
Quantity & Value & Description \\
\midrule
Initial footprint & Disconnected clover-like shape & Used for all optimization cases \\
Pretraining optimizer & Adam & Shape warm-up only \\
Pretraining iterations & $400$ & Before main optimization \\
Pretraining learning rate & $10^{-2}$ & Shape warm-up \\
\bottomrule
\end{tabular}
\end{table}

\subsection{Scalar Parameterization and Optimizer}

Scalar variables are mapped to their bounds as
\begin{equation}
    a = a_{\min} + \sigma(\eta)(a_{\max}-a_{\min}),
\end{equation}
where $\eta$ is the unconstrained trainable parameter.

\begin{table}[H]
\centering
\caption{Main optimizer settings.}
\label{tab:optimizer_settings}
\begin{tabular}{lll}
\toprule
Quantity & Value & Description \\
\midrule
Optimizer & Adam & Main optimization \\
Gradient clipping & $50$ & Global norm \\
Geometry learning rate & $10^{-3}$ & INR parameters \\
Scalar learning rate & $5\times10^{-3}$ & $T_{\mathrm{grill}}$, $t_{\mathrm{cook}}$, $t_z$, $f_f^0$ \\
Warmup & $10$ epochs & Learning-rate warmup \\
Schedule & Cosine decay & With restart period \\
Restart period & $20000$ epochs & Learning-rate cycle length \\
Minimum LR factor & $0.2$ & Cosine lower bound \\
Maximum epochs & $20000$ & Main optimization \\
\bottomrule
\end{tabular}
\end{table}

\begin{table}[H]
\centering
\caption{Trainable variables by optimization mode.}
\label{tab:trainable_by_mode}
\begin{tabular}{lll}
\toprule
Mode & Trainable variables & Frozen variables \\
\midrule
Geometry-only & $\Theta_{\mathrm{geo}}$ & $T_{\mathrm{grill}}, t_{\mathrm{cook}}, t_z, f_f^0$ \\
Joint & $\Theta_{\mathrm{geo}}, T_{\mathrm{grill}}, t_{\mathrm{cook}}, t_z, f_f^0$ & None of the scalar design variables \\
Joint, fixed thickness & $\Theta_{\mathrm{geo}}, T_{\mathrm{grill}}, t_{\mathrm{cook}}, f_f^0$ & $t_z$ \\
\bottomrule
\end{tabular}
\end{table}

\subsection{Finite-Difference Update Details}

\begin{table}[H]
\centering
\caption{Finite-difference implementation details.}
\label{tab:numerical_update_details}
\begin{tabular}{p{0.25\textwidth}p{0.68\textwidth}}
\toprule
Component & Implementation \\
\midrule
Material mask &
The material domain is represented by $M(\mathbf{x},t)=\sigma[-\alpha_\phi\phi(\mathbf{x},t)]$, with $\alpha_\phi=1000~\mathrm{m^{-1}}$. \\

Fictitious air &
The surrounding domain uses $k_{\mathrm{air}}=0.026~\mathrm{W/mK}$, with numerical heat capacity values selected for stable explicit updates. \\

Transport &
Temperature, SDF, phase fractions, and history variables are updated using upwind finite differences for advective terms. \\

Thermal conduction &
Conductive fluxes are evaluated in finite-difference flux form using harmonic averaging of $k_{\mathrm{eff}}$ at cell faces. Non-periodic boundary treatment is used to avoid wraparound fluxes. \\

Moisture diffusion &
Diffusion is applied only to the water fraction $f_w$ using harmonic averaging of $D_{\mathrm{eff}}$. Fat and protein are not diffused. \\

Surface fluxes &
Convective and plate-contact heat fluxes are implemented as surface-weighted source terms on the Eulerian grid. Plate heating is restricted to downward-facing projected contact regions. \\

Liquid loss &
Evaporation reduces the local water fraction and removes energy according to $H_{\mathrm{vap}}$. Water squeeze-out and fat drip are controlled by the calibrated coefficients in Table~\ref{tab:bc_transport_constants}. \\

Shrinkage and mass accounting &
Shrinkage advects the SDF and material fields. Protein is treated as the structural phase and renormalized after shrinkage to conserve total protein mass. \\

Flip operation &
At $t_{\mathrm{flip}}$, fields are flipped in the thickness direction and shifted so the new lower surface contacts the plate. Thermal and material histories are preserved. \\

History variables &
Maximum historical temperature, cumulative toughness exposure, and top/bottom surface tracers are tracked during the rollout. \\
\bottomrule
\end{tabular}
\end{table}

\subsection{Loss Implementation}

The smooth positive-part function is denoted by
\begin{equation}
    \mathrm{softplus}_{\alpha}(x)
    =
    \frac{1}{\alpha}\log\left(1+\exp(\alpha x)\right).
\end{equation}

\begin{table}[H]
\centering
\caption{Task-loss implementation.}
\label{tab:task_loss_implementation}
\begin{tabular}{p{0.22\textwidth}p{0.70\textwidth}}
\toprule
Loss term & Implementation \\
\midrule
Scalarization &
$\mathcal{S}(\cdot)$ is Log-Sum-Exp with $\beta=0.1$. \\
Core heating &
$\mathcal{L}_{\mathrm{core}}=0.35d_{\max}^2+2.0d_{\mathrm{mean}}^2$, where $d_{\max}$ and $d_{\mathrm{mean}}$ are smoothed deficits below $T_{\mathrm{core}}^\ast$ in the interior mask. \\
Liquid retention &
$\mathcal{L}_{\mathrm{retention}}
=
150\,\mathrm{softplus}_{25}(r_{\mathrm{ret}}^\ast-r_{\mathrm{ret}})
+
800\,\mathrm{softplus}_{100}(r_{\mathrm{ret,danger}}-r_{\mathrm{ret}})$. \\
Surface browning &
Top and bottom tracer surfaces are evaluated separately. Temperatures below $140^\circ\mathrm{C}$ are penalized as pale regions and temperatures above $195^\circ\mathrm{C}$ are penalized as burned regions. The two losses are averaged. \\
Toughness &
$\mathcal{L}_{\mathrm{tough}}
=
0.5\,\mathrm{softplus}_{2}(p_{\mathrm{ruin}}-5)
+
0.625\,\mathrm{softplus}_{2}(p_{\mathrm{ruin}}-20)$. \\
\bottomrule
\end{tabular}
\end{table}

\begin{table}[H]
\centering
\caption{Constraint and admissibility losses.}
\label{tab:constraint_admissibility_losses}
\begin{tabular}{lll}
\toprule
Term & Weight / implementation & Purpose \\
\midrule
$\mathcal{L}_{\mathrm{mass}}$ & $5000$ & Penalizes initial mass deviation \\
$\mathcal{L}_{\mathrm{wall}}$ & $4000$ inside term, then $\times100$ & Penalizes material outside allowable plate region \\
$\mathcal{L}_{\mathrm{center}}$ & $800$ & Penalizes footprint drift from plate center \\
$\mathcal{L}_{\mathrm{eik}}$ & $5$ & Preserves SDF character near the boundary \\
$\mathcal{L}_{\mathrm{Euler}}$ & $500$ & Penalizes disconnected components \\
$\mathcal{L}_{\mathrm{length}}$ & $5\times5$ differentiable morphology & Penalizes thin bridges and small holes \\
$\mathcal{L}_{\mathrm{TV}}$ & $0.1$ & Suppresses high-frequency boundary noise \\
\bottomrule
\end{tabular}
\end{table}

The topology term is implemented as
\begin{equation}
    \mathcal{L}_{\mathrm{topology}}
    =
    500\mathcal{L}_{\mathrm{Euler}}
    +
    \mathcal{L}_{\mathrm{length}}
    +
    0.1\mathcal{L}_{\mathrm{TV}}.
\end{equation}

\FloatBarrier

\end{appendices}

\end{document}